\documentclass[aps,prl,twocolumn,superscriptaddress,showpacs,amsmath,amssymb,floatfix]{revtex4-1}

\usepackage{graphicx}
\usepackage{dcolumn}
\usepackage{bm}
\usepackage{subfigure}

\begin{document}

\title{Geometrical and transport properties of sequential adsorption clusters}

\author{E. B. Ara\'ujo}
\affiliation{Departamento de F\'isica, Universidade Federal do Cear\'a, Campus do Pici 60451-970 Fortaleza, Cear\'a, Brazil}
\author{A. A. Moreira}
\affiliation{Departamento de F\'isica, Universidade Federal do Cear\'a, Campus do Pici 60451-970 Fortaleza, Cear\'a, Brazil}
\author{H. J. Herrmann}
\affiliation{Departamento de F\'isica, Universidade Federal do Cear\'a, Campus do Pici 60451-970 Fortaleza, Cear\'a, Brazil}
\affiliation{Computaional Physics for Engineering Materials, IfB, ETH Zurich, Wolfgang-Pauli-Strasse 27, CH-8093 Zurich, Switzerland}
\author{L. R. da Silva}
\affiliation{Departamento de F\'isica, Universidade Federal do Rio Grande do Norte, 59072-970 Natal, Rio Grande do Norte, Brazil}
\author{J. S. Andrade Jr.}
\affiliation{Departamento de F\'isica, Universidade Federal do Cear\'a, Campus do Pici 60451-970 Fortaleza, Cear\'a, Brazil}

\date{\today}

\begin{abstract}
We investigate transport properties of percolating clusters generated by irreversible
cooperative sequential adsorption (CSA) on square lattices with Arrhenius rates
given by $k_{i}\equiv q^{n_{i}}$, where $n_{i}$ is the number of occupied neighbors
of the site $i$, and $q$ a controlling parameter. Our results show
a dependence of the prefactors
on $q$ and a strong finite size effect for small values of this parameter, both
impacting the size of the backbone and the global conductance of the system. These
results might be pertinent to practical applications in processes involving adsorption
of particles.
\end{abstract}

\maketitle

\section{Introduction}
Transport properties in disordered media have been successfully and extensively
studied within percolation theory
\cite{soares1992,isi,Stauffer,pre51,pre55,soares1999,soares1,soares2,pose}.
Related to percolation is random sequential adsorption RSA of particles on a
lattice \cite{evans1993random}. Sites are occupied sequentially independently of
other sites and the system exhibits a spanning cluster at the percolation threshold.

Despite the accomplishments of the RSA model in describing disordered media \cite{evans1993random,nuno1},
interactions constitute an important feature in real physical systems.
A common phase observed in adsorption is the $c(2\times2)$
\cite{jager,chang1987formation,evans1987nonequilibrium},
in which particles are
arranged in a checkerboard pattern \cite{wood1964vocabulary}, showing that nearest
neighbor (NN) exclusion plays
an important role in this process. This correlated filling has been studied 
for a chain molecule with reactive sites \cite{keller1982} and on square
lattices, using both hierarchical rate equations \cite{evans1983irreversible}
and percolation theory \cite{evans1988},
called in this case cooperative sequential adsorption (CSA). Filling of random and correlated
`lattice animals' has been the subject of an extensive review \cite{evans1993random}.
The fractal dimension of the spanning cluster on CSA
models has already been shown to be the same for traditional percolation
\cite{evans1988,scott,nuno2}.
However, the island formation process is known to be
dependent on the filling ratios \cite{evans1988nonequilibrium,sanders1988correlated},
and produces clusters with peculiar shapes for different ratios. Here we show and
quantify the dependence of the transport properties of the spanning cluster on
different ratios.

\begin{figure}
\includegraphics*[width=\columnwidth]{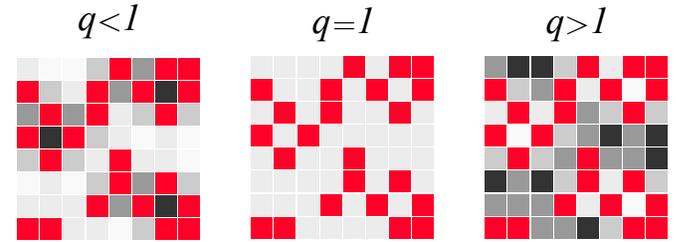}
\caption{\label{fig:sample}Representation of the CSA process on a square lattice
for non interacting ($q=1$), repulsive ($q<1$) and attractive ($q>1$) particles
with the same occupied sites (in red). Empty sites are shown in grayscale.
The probability of occupying an empty site increases with its lightness.
For non interacting particles, the probability of occupying any empty
site is the same, regardless the number of occupied nearest neighbors.
For interacting particles it depends on the number of occupied NN. The difference
between probabilities depends on the strength of the interaction, regulated
by the parameter $q$, and the configuration of the system.}
\end{figure}

In the CSA model, a particle irreversibly occupies an empty site $i$ with a (multiplicative) rate
given by
\begin{equation}
k_{i} \equiv q^{n_{i}},
\end{equation}
where $n_{i}$ is the number of occupied nearest neighbors of site $i$.
The parameter $q$ controls the strength an occupied site influences the filling
of a neighboring unoccupied site: for $q>1$ there is an attractive filling and
for $q<1$ a repulsive filling. For $q=1$, every unoccupied site can be filled with
the same probability, the model being equivalent to RSA.

\begin{figure}[b]
\includegraphics[width=0.8\columnwidth]{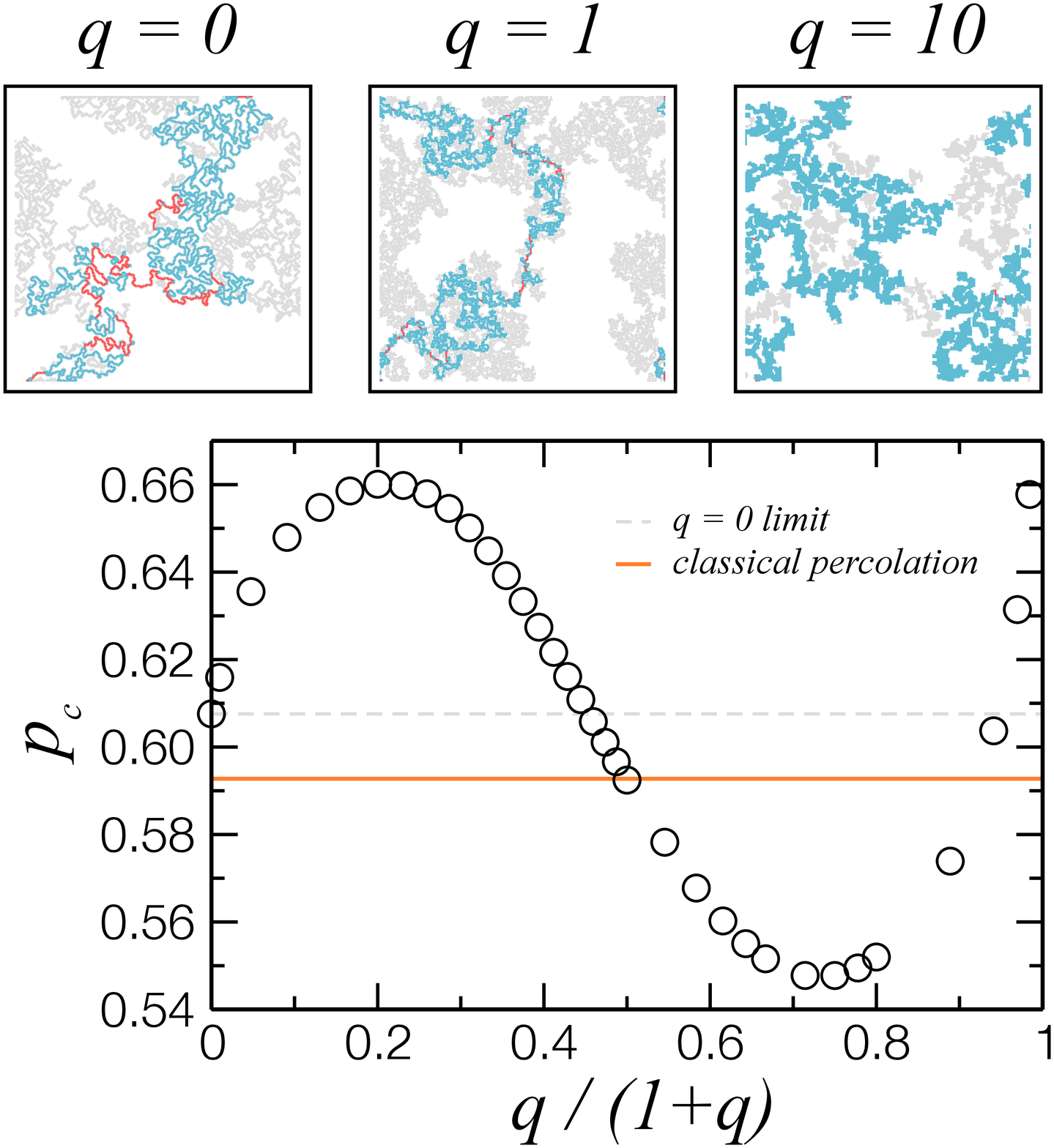}
\caption{\label{fig:clusters}Top: typical spanning clusters
on the onset of percolation for $q=0$, $q=1$ and $q=10$. The spanning cluster
appears in gray, the conducting backbone in blue and cutting bonds in red. For
larger $q$, the fraction of sites of the spanning cluster present in the backbone
is closer to 1. The number of cutting bonds generally decreases with $q$.
Bottom: dependence of the percolation threshold $p_{c}$ as
a function of the parameter $q$. The orange line is the value for classical
percolation, with $q/(1+q)=0.5$. The dashed gray line is the value for $q=0$. Average
values are obtained for $L=4096$ and calculated over 2000 realizations.}
\end{figure}

\section{Method}
We implement this model on a square lattice of size $L$ with periodic boundary
conditions in horizontal direction. At each iteration step, we fill a random unoccupied
site $i$ chosen with probability
\begin{equation}
\label{eq:prob-choose}
P_{i} = \frac{q^{n_i}}{\sum_{n_{j}}T_{n_{j}}q^{n_{j}}}
\end{equation}
where $T_{n_{j}}$ is the total number of sites with $n_{j}$ NN and $q\neq0$.
Fig. \ref{fig:sample} illustrates the model for several values of $q$. For the
special case of $q=0$, we notice that in the early stages of the process for very
low $q$
\begin{equation}
\lim_{q\rightarrow0}T_{0}P(0) = 1.
\end{equation}
Hence, only sites with unoccupied NN would be filled and the system would never
percolate. Filling would occur as in RSA with NN exclusion and stop at a jammed
state with an occupation fraction of 0.365 \cite{evans1983irreversible,Meakin,dwyer1977study,kertesz1982threshold,baram1989dynamics,fan1991use,dickman1991random}.
We modify the model for the special case of $q\rightarrow 0^{+}$, labeled
henceforth $q=0$, to obtain results compatible with this limit in the following
way. Sites are filled with NN exclusion until the system reaches the
jammed state. Then all clusters in the system have size one, but no spanning
cluster is formed yet. Next, sites with only one neighbor are then
allowed to be occupied. When these sites are extinguished, the ones with two
neighbors are then occupied and so on, until a spanning cluster appears, promoting
the global connection of the system.

Once the spanning cluster is obtained, the backbone and the cutting bonds are identified
using the burning method \cite{Hans}. To calculate the conductivity, we assume that
there is a resistor of conductance $g_{ij}$ between every pair of sites $i$
and $j$ of the system. The value of the conductance is
\begin{equation}
g_{ij}=
	\begin{cases}
		1, &\text{if $i \wedge j \in$ backbone;}\\
		0, &\text{otherwise.}
	\end{cases}
\end{equation}
An arbitrary current is then applied between two sites ($i=1$ and $i=N_{back}$,
the number of sites in the backbone)
in opposite sides of the spanning cluster and Kirchhoff's current law is imposed
at each site. As a result, we obtain a set of coupled linear algebraic equations:
\begin{equation}
\label{eq:system}
\sum_{i \neq j} M_{ij} (V_{i} - V_{j}) =
	\begin{cases}
		-1, &\text{if $i=1$;}\\
		1, &\text{if $i=N_{back}$;}\\
		0, &\text{otherwise}.
	\end{cases}
\end{equation}
where $M$ is the Laplacian matrix of the backbone sites. $M_{ij}$ is given by
\begin{equation}
M_{ij} =
	\begin{cases}
		n_{i}, &\text{if $i=j$;}\\
		-1, &\text{if $i$ and $j$ are neighbors;}\\
		0, &\text{otherwise}.
	\end{cases}
\end{equation}
We solve the system of Eqs. \ref{eq:system}, for $i,j=2,3,\dots,N_{clus}$ to obtain
the voltages in each resistor and therefore the global conductance of the system.
For each quantity an average is obtained over at least 2000 realizations for a
given value of $L$.

\begin{figure}[t]
\includegraphics*[width=0.9\columnwidth]{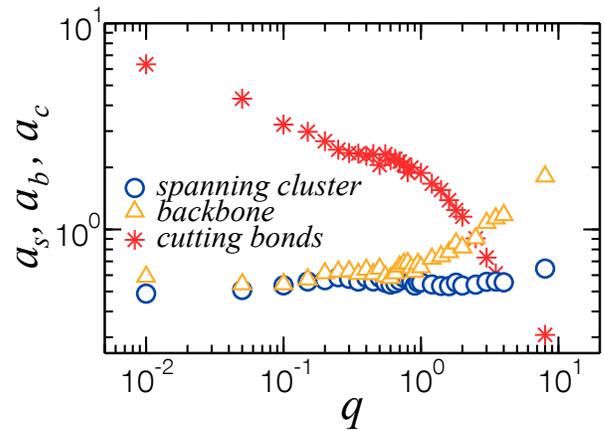}
\caption{\label{fig:coef}Dependence on the parameter $q$ of the prefactor in the
finite-size scaling laws for the masses of spanning cluster, $M_{s}=a_{s}L^{d_{s}}$,
backbone, $M_{b}=a_{b}L^{d_{b}}$, and cutting bonds, $M_{c}=a_{c}L^{d_{c}}$. The
cutting bonds prefactor $a_{c}$ decreases
monotonically with $q$ but faster for attractive filling ($q>1$). $a_{s}$ is
close to $a_{b}$ for repulsive filling, but for attractive filling it increases faster.}
\end{figure}

\section{Results}

The dependence on $q$ of the percolation threshold \cite{sanders1988correlated}
is calculated to high precision for system sizes up to $L=4096$, as shown in Fig.
\ref{fig:clusters}. Also, typical realizations of the system at the critical
point are shown for three distinct values of $q$. Previous studies indicate that
this model belongs to the same universality class of traditional percolation \cite{evans1988}.
Our data and analysis confirm that the fractal dimensions of the masses of the
cluster, backbone, cutting bonds and of the conductivity are indeed the same as in
traditional percolation.

Nevertheless, there is a clear difference in shape for the clusters obtained
for different values of $q$ \cite{evans1988}. For very small values of $q$, the
cluster resembles a warped wire and presents a large number of cutting bonds.
As $q$ increases and the repelling force becomes weaker (and turns to attraction
for $q>1$), the clusters become more massive. At the same time more sites are present
in the backbone and there are less cutting bonds present.

\begin{figure}[b]
\includegraphics*[width=\columnwidth]{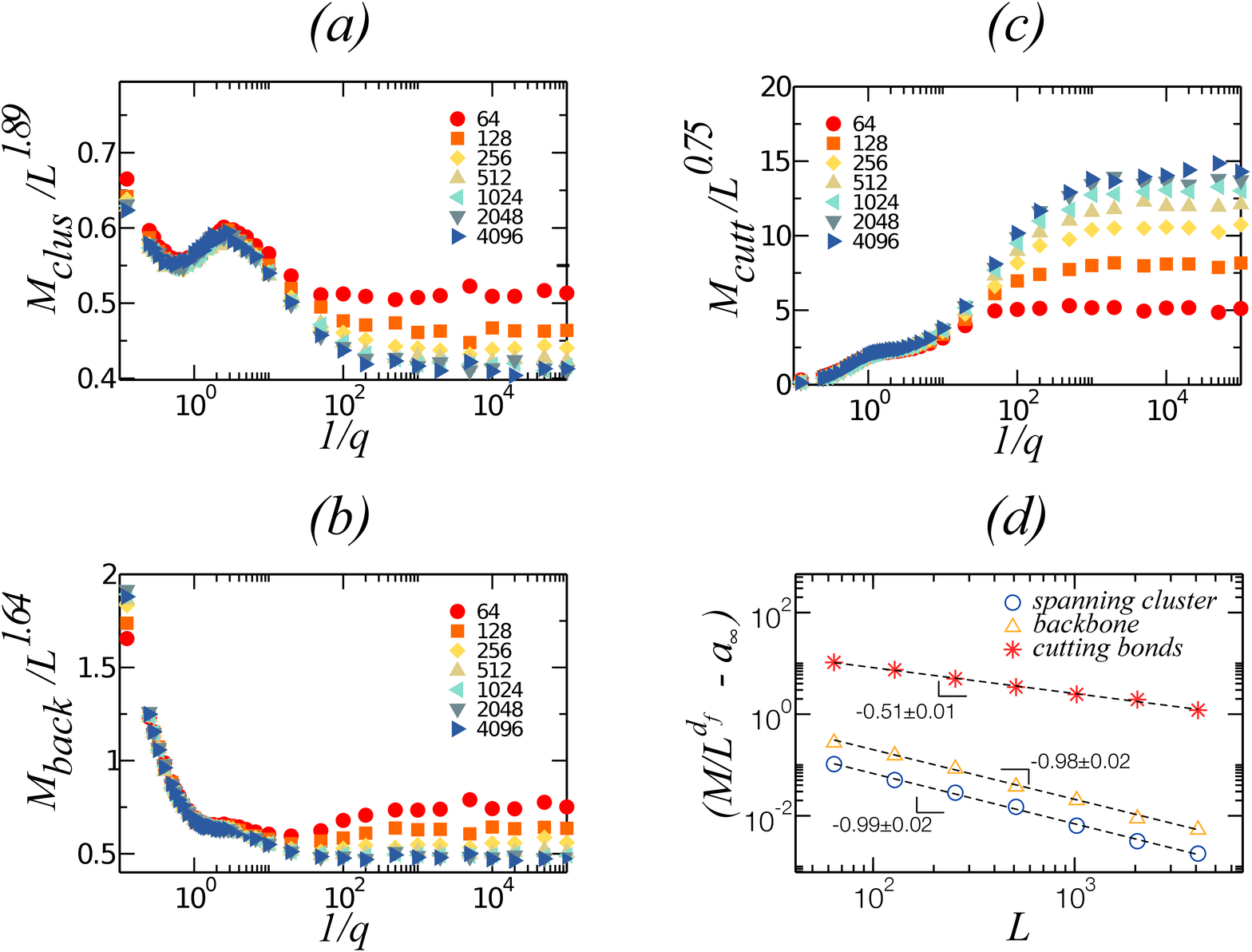}
\caption{\label{fig:broom}$(a)-(c)$Effect of the finite size of the samples on the scaling
behavior of masses of the spanning cluster, backbone and cutting bonds. We show the
mass divided by the system size to the power of the corresponding fractal dimension for different
$q$. For small $q$, the convergence towards the thermodynamic limit becomes very slow.
$(d)$ Convergence of the $a_{\infty}$ towards the thermodynamic limit for $q\rightarrow 0$.
We calculated $a_{\infty}$ from the best power law fit of $(M/L^{d_{f}}-a_{\infty})$ for each quantity.}
\end{figure}

We measure this effect calculating the prefactors from power law fits at criticality
for each mass with system size, $M_{s}=a_{s}L^{d_{s}}$,
backbone, $M_{b}=a_{b}L^{d_{b}}$, and  $M_{c}=a_{c}L^{d_{c}}$, where $d_s$, $d_b$
and $d_c$ are the corresponding critical exponents. The results are shown in Fig.
\ref{fig:coef}. While the spanning cluster mass prefactor, $a_{s}$, varies little
with $q$, the cutting bonds
mass prefactor $a_{c}$ decreases with $q$. For repulsive filling ($q<1$) the
backbone mass prefactor $a_{b}$ is close to $a_{m}$ but increases faster for
attractive filling ($q>1$). As a result, more sites of the spanning
cluster are present in the backbone.
These quantities, however, deviate from the expected power law behavior depending
on the value of $q$. For $q\ll1$ there are strong finite size effects, as shown
in Fig. \ref{fig:broom}. We plot the masses of the spanning cluster, backbone and cutting
bonds divided by the system size to the power of their corresponding fractal dimension
(the prefactor for that system size) as function of $q$. For small $q$, the convergence towards
the thermodynamic limit becomes very slow. We show that the low $q$ prefactor
converges to a finite limit $a_{\infty}$ as a power law of the system size. These
values are shown in Table I.

\begin{table}
\caption{Values of the low $q$ prefactor $a_{\infty}$ for which the curves in Fig.
\ref{fig:broom} converge, calculated by finding the best fit of $(M/L^{d_{f}}-a_{\infty})$
using a power law function. The exponent of the power law, $\alpha$, is also
given.}
{\begin{tabular}{l c c}
\hline
Low $q$ Prefactor & $a_{\infty}$ & $\alpha$ \\
\hline
\hline
Spanning cluster mass & $0.4104$ & $-0.99 \pm0.02$\\
Backbone mass & $0.4708$ & $-0.98\pm0.02$\\
Cutting bonds mass & $15.51$ & $-0.51\pm0.01$\\
\hline
\end{tabular}
\label{table:limits}}
\end{table}

\begin{figure}
\includegraphics*[width=0.9\columnwidth]{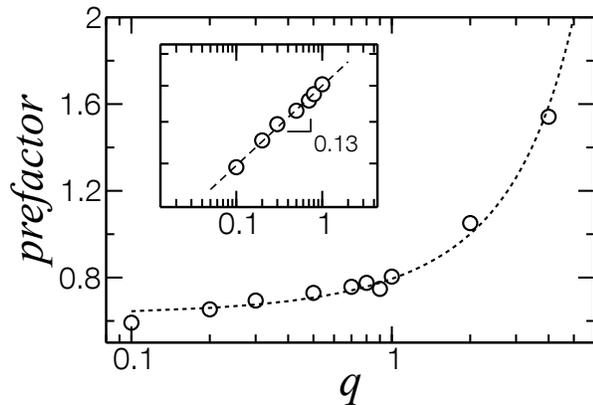}
\caption{\label{fig:cond}Behavior of the prefactor $a$ of the conductance
as a function of $q$. The calculated prefactors increase monotonically with $q$
in this range, showing that clusters formed by repulsive particles are
particularly poor conductors as compared to ones formed by attractive particles.
The dashed line is an exponential guide for the eyes. In the inset is shown the behavior
of the prefactor as $q$ approaches 0. The dashed line is a power law with exponent
0.13. Each point is an average over 5000 realizations. The error bars are smaller
than the symbols.}
\end{figure}

The results for the prefactor of the global electrical conductance are shown in Fig.
\ref{fig:cond} for $q \in [0.1,4]$. In this range it increases monotonically with
$q$. For repulsive filling it increases slowly, being well fitted by a power
law with exponent 0.13. For $q=4$ the global conductance prefactor is more than
$50\%$ above the value of traditional percolation. This indicates that the more
attractive the particles are, the better conductor is the conducting backbone
cluster.

\section{Conclusions}

In summary, we have studied transport properties of percolating clusters generated
by the CSA model with nearest neighbors correlations given by multiplicative rates
modulated by a parameter $q$. Our results confirm that the mass of the backbone
and the number of cutting bonds as well as the global conductance scale as power
laws with the system size with the same exponents as in traditional percolation.
We have shown that changes in the parameter $q$ result in changes in all prefactors
of these properties. These changes affect the spanning
cluster shape, the coverage of the backbone, the number of cutting bonds and
the global conductance of the system.
As $q$ becomes large, i.e., within the very attractive interaction regime, occupied
sites are more likely to have occupied NN. This increases the fraction of sites
in the backbone, allowing alternative paths for the current to flow in the cluster
thus increasing the conductance of the system. Finally, we have shown the strong finite
size effect for $q\ll1$ and calculated the limiting values of the corresponding
power-law prefactors.

\section*{Acknowledgments}
We acknowledge financial support from the Brazilian agencies CNPq, CAPES,
FUNCAP and European Research Council (ERC) Advanced Grant 319968-FlowCCS.

\end{document}